\documentclass[10.75pt, a4paper]{article}
\usepackage{titlesec}
\titleformat{\section}
  {\bf\sffamily}
  {\thesection. }
  {5pt}
  {\MakeUppercase}
\renewcommand{\thesection}{\Roman{section}} 

\titleformat{\subsection}
  {\bf\sffamily}
  {\thesubsection. }
  {5pt}{}
  
\renewcommand{\thesubsection}{\Alph{subsection}}

\usepackage[affil-it]{authblk} 
\usepackage{etoolbox}
\usepackage{lmodern}

\usepackage{eurosym}

\usepackage[utf8]{inputenc}
\usepackage{geometry}
\usepackage[hidelinks]{hyperref}
\usepackage[citestyle=numeric,style=phys,biblabel=brackets,backend=bibtex,sorting=none,url=false,doi=false, natbib]{biblatex}
\bibliography{bibliography}
\DefineBibliographyStrings{english}{andothers={\itshape et\addabbrvspace al\adddot}}

\usepackage{csquotes}
\usepackage[]{authblk} 
\usepackage{etoolbox}
\usepackage{lmodern}

\usepackage{amsmath}
\usepackage{enumerate}
\usepackage{enumitem}
\usepackage{graphicx}
\usepackage{siunitx}
\usepackage{float}
\usepackage[]{parskip} 

\makeatletter
\patchcmd{\@maketitle}{\LARGE \@title}{\fontsize{16}{19.2}\selectfont\@title}{}{}
\makeatother

\usepackage[normalem]{ulem}

\usepackage{graphicx}
\usepackage{dcolumn}
\usepackage{bm}
\usepackage{tikz}
\usetikzlibrary{math}
\usetikzlibrary{shapes.geometric, arrows}
\usetikzlibrary{positioning}
\usetikzlibrary{shapes,arrows}
\usetikzlibrary{intersections}
\usepackage{csvsimple}
\usepackage{changepage}
\usepackage[tbtags]{mathtools}
\usepackage{siunitx}
\usepackage{csquotes}

\usepackage{float}
\usepackage{subfigure}
\usepackage[
	nonumberlist, 				
	acronym,      				
	nomain,						
	nopostdot,					
]{glossaries}
\usepackage{glossary-superragged}
\usepackage[hidelinks]{hyperref}
\usepackage{color, colortbl}

\usepackage{wrapfig}

\usepackage{pgfplots}
\usepackage{tikz}
\usetikzlibrary{arrows}
\usepackage{amsmath}
\pgfplotsset{compat=newest}
\usepgfplotslibrary{fillbetween}
\usetikzlibrary{calc}
\def\centerarc[#1](#2)(#3:#4:#5)
{ \draw[#1] ($(#2)+({#5*cos(#3)},{#5*sin(#3)})$) arc (#3:#4:#5); }

\usepackage{amssymb}

\usepackage{nicefrac}

\usepackage{abstract}

\usepackage{multirow}
\usepackage{tabularx}
\usepackage{booktabs}
\usepackage{array}
\usepackage{longtable}
\newcolumntype{L}[1]{>{\raggedright\let\newline\\\arraybackslash\hspace{0pt}}m{#1}}
\newcolumntype{C}[1]{>{\centering\let\newline\\\arraybackslash\hspace{0pt}}m{#1}}
\newcolumntype{R}[1]{>{\raggedleft\let\newline\\\arraybackslash\hspace{0pt}}m{#1}}

\newacronym{3d}{3D}{three dimensional}
\newacronym{am}{AM}{additive manufacturing}
\newacronym{fdm}{FDM}{fused deposition modeling}
\newacronym{ism}{ISM}{in-space manufacturing}
\newacronym{iss}{ISS}{International Space Station}
\newacronym{fcb}{FCB}{Functional Cargo Block}
\newacronym{dem}{DEM}{discrete element method}
\newacronym{md}{MD}{molecular dynamics}
\newacronym{dc}{DC}{direct-current}
\newacronym[plural=PFCs,firstplural=parabolic flight campaigns (PFCs)]{pfc}{PFC}{Parabolic Flight Campaign}
\newacronym{fft}{FFT}{Fast Fourrier Transform}
\newacronym{cad}{CAD}{Computer Assisted Design}
\newacronym{ptfe}{PTFE}{polytetrafluoroethylene}
\newacronym{ps}{PS}{polystyrene}
\newacronym{nasa}{NASA}{National Aeronautics and Space Administration}
\newacronym{esamm}{ESAMM}{Extended Structure Additive Manufacturing Machine}
\newacronym{amf}{AMF}{Additive Manufacturing Facility}
\newacronym{us}{US}{United States}
\newacronym{usa}{USA}{United States of America}
\newacronym{bmgs}{BMGs}{Bulk Metallic Glasses}
\newacronym{esa}{ESA}{European Space Agency}
\newacronym{si}{SI}{International System of Units, abbreviated from French \textit{Syst\`{e}me International (d'unit\'{e}s)}}
\newacronym{dlr}{DLR}{German Aerospace Center}
\newacronym{liggghts}{LIGGGHTS}{\acrshort{lammps} Improved for General Granular and Granular Heat Transfer Simulations}
\newacronym{lammps}{LAMMPS}{Large-scale Atomic/Molecular Massively Parallel Simulator}
\newacronym{sjkr}{SJKR}{Simplified Johnson-Kendall-Roberts}
\newacronym{ded}{DED}{Directed Energy Deposition}
\newacronym{slm}{SLM}{Selective Laser Melting}
\newacronym{sls}{SLS}{Selective Laser Sintering}
\newacronym{eva}{EVA}{Extra-Vehicular Activity}
\newacronym{sem}{SEM}{Scanning Electron Microscopy}
\newacronym{RPM}{RPM}{Ramdom Positioning Machine}
\newacronym{rpm}{rpm}{revolutions per minute}
\newacronym{rise}{RISE}{Research Internships in Science and Engineering}
\newacronym{daad}{DAAD}{German Academic Exchange Service, abbreviated from German \textit{Deutscher Akademischer Austauschdienst}}
\newacronym{fsm}{FSM}{finite-state machine}
\newacronym{ir}{IR}{infrared}
\newacronym{pcbs}{PCBs}{Printed Circuit Boards}
\newacronym{pcb}{PCB}{Printed Circuit Board}
\newacronym{mcr}{MCR}{Modular Compact Rheometer}
\newacronym{sff}{SFF}{Solid Freeform Fabrication}
\newacronym{uv}{UV}{ultraviolet}
\newacronym{abs}{ABS}{acrylonitrile butadiene styrene}
\newacronym{hpde}{HPDE}{high density polyethylene}
\newacronym{pei}{PEI}{polyetherimide}
\newacronym{bff}{BFF}{BioFabrication Facility}
\newacronym{lens}{LENS}{Laser Engineered Net Shaping}
\newacronym{cnc}{CNC}{Computer Numerical Control}
\newacronym{ebf3}{EBF$^3$}{Electron Beam Free-Form Fabrication}
\newacronym{leo}{LEO}{Low Earth Orbit}
\newacronym{pc}{PC}{polycarbonate}
\newacronym{crissp}{CRISSP}{Customisable Recyclable International Space Station Packaging}
\newacronym{Athena}{Athena}{Advanced Telescope for High-ENergy Astrophysics}
\newacronym{lbm}{LBM}{Laser Beam Melting}
\newacronym{bam}{BAM}{Federal Institute for Materials Research and Testing, abbreviated from German \textit{Bundesanstalt f\"{u}r Materialforschung und-pr\"{u}fung}}
\newacronym{pbf}{PBF}{powder bed fusion}
\newacronym{eb}{EB}{Electron Beam}
\newacronym{2d}{2D}{two dimensional}
\newacronym{4d}{4D}{four dimensional}
\newacronym{ft4}{FT4}{Freeman Technology 4 Powder Rheometer}
\newacronym{dsc}{DSC}{Differential Scanning Calorimetry}
\newacronym{pmma}{PMMA}{polymethylmethacrylate}
\newacronym{1g}{$1g$}{gravity on-ground}
\newacronym{mug}{$\mu g$}{microgravity}
\newacronym{bcm}{BCM}{Box Counting Method}
\newacronym{mct}{MCT}{Mode Coupling Theory}
\newacronym{gmct}{gMCT}{granular Mode Coupling Theory}
\newacronym{itt}{ITT}{Integration Through Transients}
\newacronym{mfc}{MFC}{Mass Flow Controller}
\newacronym{ct}{CT}{computed tomography}
\newacronym{xct}{XCT}{X-ray computed tomography}
\newacronym{cv}{CV}{curriculum vitae}
\newacronym{pi}{PI}{principal investigator}
\newacronym{osp}{OSP}{orthogonal superimposed perturbation}
\newacronym{npi}{NPI}{Network Partnering Initiative}
\newacronym{ecsat}{ECSAT}{European Centre for Space Applications and Telecommunications}
\newacronym{eac}{EAC}{European Astronaut Centre}
\newacronym{estec}{ESTEC}{European Space Research and Technology Centre}
\newacronym{fps}{fps}{frames per second}
\newacronym{pdf}{pdf}{probability density function}
\newacronym{al}{Al}{aluminium}
\newacronym{ss}{\textit{SS}}{\textit{Smooth Surface}}
\newacronym{rs}{\textit{RS}}{\textit{Rough Surface}}
\newacronym{rcp}{rcp}{random close packing}
\newacronym{iop}{IoP UvA}{Institute of Physics of the University of Amsterdam}
\newacronym{mp}{MP}{Institute of Material Physics for Space}
\newacronym{elgra}{ELGRA}{European Low Gravity Research Association}
\newacronym{zarm}{ZARM}{Center of Applied Space Technology and Microgravity}
\newacronym{piv}{PIV}{particle image velocimetry}
\usepackage[framemethod=tikz]{mdframed}
\usepackage{lipsum}
\usepackage{dirtytalk}
\usepackage{tcolorbox}
\usepackage[font=footnotesize]{caption}
\newtcolorbox{mybox}[1]{colback=green!6!white,colframe=black!75!black,fonttitle=\bfseries,title=#1}
\newtcolorbox{mybox2}{colback=red!5!white,colframe=red!75!black}

\usepackage{pifont}

\usepackage{soul,xcolor}
\setstcolor{red}

\usepackage{layouts}

\usepackage{xcolor,hyperref}
\hypersetup{
   colorlinks,
   linkcolor={blue!50!black},
   citecolor={blue!50!black},
   urlcolor={blue!80!black}
} 

\definecolor{mycolor}{rgb}{0.122, 0.435, 0.698}

 \title{Laser-induced droplet deformation: \\
 curvature inversion explained from instantaneous pressure impulse} 
\author[1,2]{Hugo França\footnote{h.franca@arcnl.nl}}
\author[1,3]{Hermann Karl Schubert}
\author[1,3]{Oscar Versolato}
\author[2]{Maziyar Jalaal}

\affil[1]{Advanced Research Center for Nanolithography, Science Park 106, 1098 XG Amsterdam, The Netherlands}
\affil[2]{Van der Waals-Zeeman Institute, Institute of Physics, University of Amsterdam, Science Park 904, Amsterdam, 1098XH, The Netherlands}
\affil[2]{Department of Physics and Astronomy, and LaserLaB, Vrije Universiteit Amsterdam, De Boelelaan 1081, 1081 HV Amsterdam, The Netherlands}

\begin{document}
\definecolor{brickred}{rgb}{0.8, 0.25, 0.33}
\definecolor{darkorange}{rgb}{1.0, 0.55, 0.0}
\definecolor{persiangreen}{rgb}{0.0, 0.65, 0.58}
\definecolor{persianindigo}{rgb}{0.2, 0.07, 0.48}
\definecolor{cadet}{rgb}{0.33, 0.41, 0.47}
\definecolor{turquoisegreen}{rgb}{0.63, 0.84, 0.71}
\definecolor{sandybrown}{rgb}{0.96, 0.64, 0.38}
\definecolor{blueblue}{rgb}{0.0, 0.2, 0.6}
\definecolor{ballblue}{rgb}{0.13, 0.67, 0.8}
\definecolor{greengreen}{rgb}{0.0, 0.5, 0.0}
\begingroup
\sffamily
\date{}
\maketitle
\endgroup

\newcommand{\Ubold}{\textbf{u}}
\newcommand{\del}[2]{\frac{\partial#1}{\partial#2}}

\begin{abstract}

{We investigate the shape of a tin sheet formed from a droplet struck by a nanosecond laser pulse. Specifically, we examine the dynamics of the process as a function of laser beam properties, focusing on the outstanding riddle of curvature inversion: tin sheets produced in experiments and state-of-the-art extreme ultraviolet (EUV) nanolithography light sources curve in a direction opposite to previous theoretical predictions. We resolve this discrepancy by combining direct numerical simulations with experimental data, demonstrating that curvature inversion can be explained by an instantaneous pressure impulse with low kurtosis. Specifically, we parametrize a dimensionless pressure width, \( W \), using a raised cosine function and successfully reproduce the experimentally observed curvature over a wide range of laser-to-droplet diameter ratios, \( 0.3 < d/D_0 < 0.8 \). The simulation process described in this work has applications in the EUV nanolithography industry, where a laser pulse deforms a droplet into a sheet, which is subsequently ionized by a second pulse to produce EUV-emitting plasma.
}


\textbf{keywords: droplet deformation $|$ capillary flows $|$ laser-fluid interaction} 

\end{abstract}

\section{Introduction}

The hydrodynamic deformation of a droplet driven by a laser pulse is a process of high interest for the nanolithography industry, in which the state-of-the-art technique for producing extreme ultraviolet (EUV) light involves deforming a tin droplet into a thin sheet which is subsequently ablated into a light-emitting plasma \cite{Bakshi2009-Book, Versolato2019, Sizyuk2020, Fomenkov2017}. Similar deformation processes can also be found in scenarios without a laser, in which the deformation is triggered by the droplet impact onto a large solid surface \cite{Bergeron2000, Josserand2016, Scheller1995} or a solid pillar \cite{Wang2017, Villermaux2011, Rozhkov2002}, for example.  In these cases, the droplet is also expanded into a thin sheet, which eventually can fragment or retract due to the effects of surface tension. These processes are also industrially relevant in a wide range of applications, such as spray coating and printing \cite{Lohse2022, Derby2010}. Hence, understanding the fluid mechanics in this deformation process is essential for further optimizing current systems.

In a typical nanolithography scenario, a spherical tin droplet is irradiated by a nanosecond laser pulse. This interaction ablates part of the droplet surface, which becomes a tin plasma that rapidly expands also on the laser pulse timescale ($\tau_p \sim 10\,$ns). As this plasma expands, it transfers momentum to the liquid droplet, which propels with a velocity $U_z \sim 100\,\text{m/s}$ and expands radially with velocity $\dot{R}_0 \sim 100\,\text{m/s}$. Therefore, this expansion happens hydrodynamically on the inertial timescale  ($\tau_i = R_0 / \dot{R}_0 \sim 100\,$ns), with $R_0 \sim 10\,\mu$m being the droplet radius. In the third stage, when the droplet is already significantly deformed into a sheet, surface tension slows down its expansion, eventually forming a rim, fragmentation, and sheet retraction. This final stage happens on the capillary timescale $\tau_c = \sqrt{\rho R_0^3/\gamma} \sim 10\,\mu$s, where $\rho = 6900\,\text{kg/m}^3$ is the liquid tin density and $\gamma = 0.55\,$N/m its surface tension \cite{Liu2020}. 

\begin{figure}[htbp]
\centering
\includegraphics[width=0.9\textwidth]{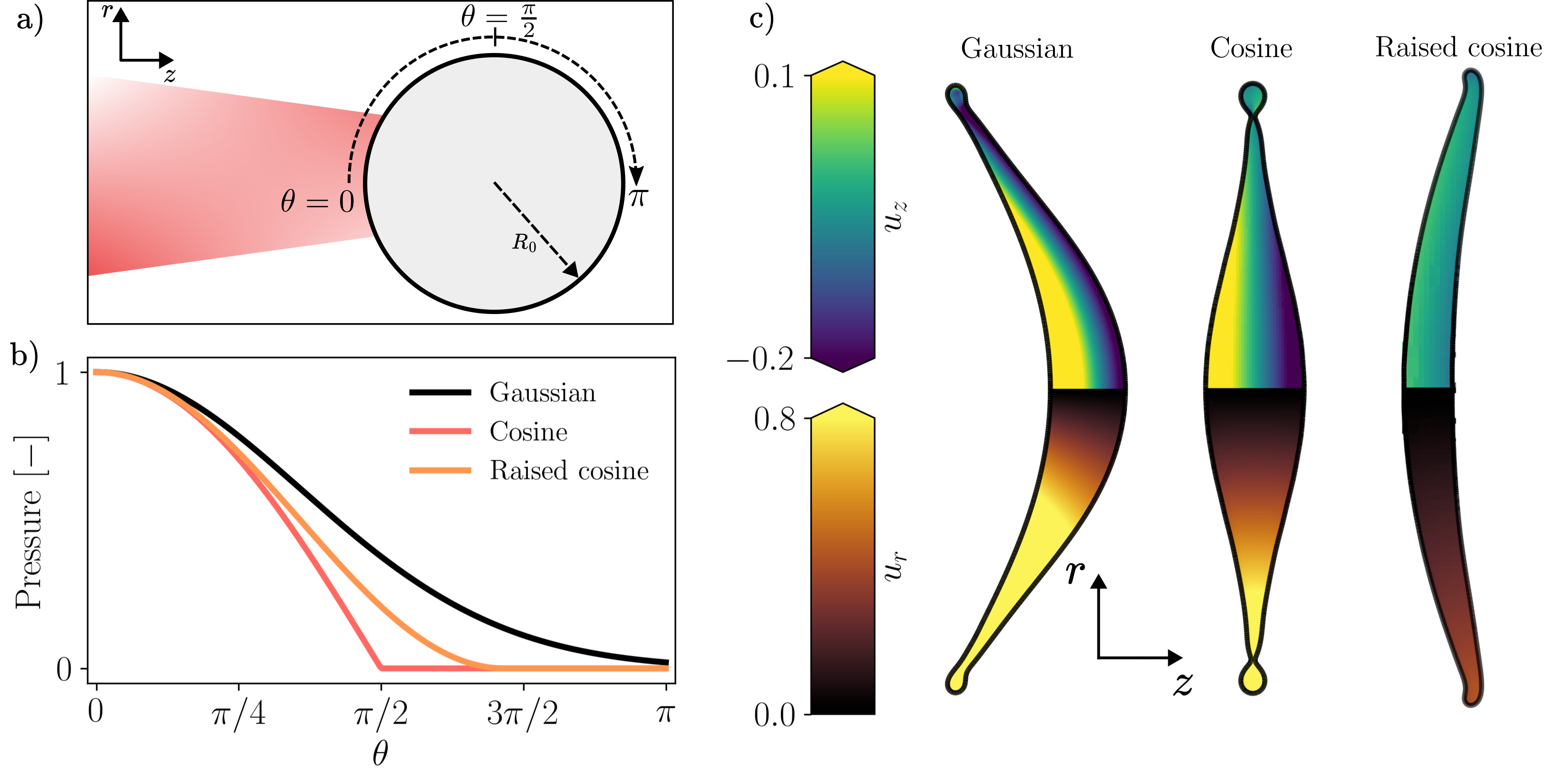}
\caption{Initialization strategy to obtain a velocity field from a pressure profile given on the droplet surface. Panel a) sketch of the droplet and laser. Panel b) examples of possible initial pressure profile functions on the droplet surface. Panel c) Radial and axial velocity components on three sheets obtained from different pressure profile types.}
\label{fig:initialization_strategy}
\end{figure}

Previous works \cite{Gelderblom2016, Klein2015} have numerically simulated laser-induced droplet deformation by assuming that the laser (and the generated plasma) interact with the droplet by applying a near-instantaneous pressure on its surface, setting it in incompressible motion. As illustrated in Figure \ref{fig:initialization_strategy}, this pressure is commonly approximated by a function $f(\theta)$ defined over the parametrized surface of the droplet, where $\theta = 0$ and $\theta=\pi$ correspond to the side of the droplet hit by the laser and the opposite side, respectively.
Different profiles have been suggested for this pressure function, most commonly a Gaussian profile is used, in which the standard deviation $\sigma$ is tuned to mimic variations in the focus size of the laser beam. A truncated cosine has also been suggested in \cite{Gelderblom2016}, and the authors show that this is the only profile capable of generating a perfectly symmetric sheet.

In  \cite{HernandezRueda2022, kurilovich_power-law_2018}, the authors utilize the full radiation-hydrodynamic code RALEF-2D to simulate the interaction between a laser and the tin, leading to the plasma formation that transfers momentum to the droplet. While their code could not simulate late-time droplet deformation, the early-time interaction between laser and tin was reported, and the authors observed pressure profiles that can match Gaussian functions in some aspects, such as propulsion-to-deformation energy ratios. However, no attention is given to the actual morphology of the deformed sheet coming from these pressure profiles. 

While Gaussians and the truncated cosine profiles have been successfully used for the purpose of these previous studies, they were not capable of reproducing one striking feature of some experiments: a sheet that curves opposite the direction of the incident laser beam, as also illustrated in figure \ref{fig:initialization_strategy}. This sheet morphology, which we refer to as a ``positive curvature'', is intriguing as it is characteristic, for example, of a droplet being pushed by a constant wind tunnel, which is not the case in a near-instantaneous focused laser interaction. Depending on the experiment's parameters, a ``negative'' curvature can also be observed. This naturally leads to the following question: is it possible to obtain a single functional form for the pressure profile to explain positive and negative curvatures, as we see both in experiments? In this work, we will go beyond the early-time dynamics and show that this is possible by suggesting a pressure profile based on a raised cosine function. As will be demonstrated later, this profile was chosen due to its capability of mimicking laser-beam focus as the Gaussian profile and providing a forward curvature that is more comparable with experiments. Correctly predicting the morphology and curvature of the sheet provides important information for EUV-light generation applications. The shape and thickness of the sheet will directly influence how much energy is necessary to ablate the liquid and generate light, such that good predictions of the sheet morphology will allow for better tuning for the optimal experimental parameters. It has also been shown in \cite{Engels2023} that the vaporization of a thin sheet creates a vapor with a shape that follows the local curvature of the sheet, indicating that controlling the morphology of the sheet is also essential for shaping the vaporization (and plasma creation) process.

The raised cosine profile is illustrated in figure \ref{fig:initialization_strategy}, along with the more classical Gaussian and cosine profiles. We also show examples of typical sheet shapes that can be numerically obtained using the three pressure profiles discussed in this paper. In the Gaussian example, a sheet that curves towards the laser is shown (negative curvature). In the raised cosine example, the sheet curves away from the laser origin (positive curvature). The center image in figure \ref{fig:initialization_strategy}c shows the truncated cosine case which presents a symmetric sheet and, according to our definition, presents zero curvature. The expressions for the three profile types are shown in table \ref{table:pressure_profiles}. Similarly to the Gaussian function, the proposed raised cosine also has a tuneable parameter $W$, which controls the width of the profile.

While in this work, we will obtain positive curvatures by using raised cosine pressure profiles, we note that this is not the only function capable of giving this sheet morphology. Other functions can also be chosen or carefully constructed to obtain similar deformation dynamics. During an exploratory study of different pressure profiles, we observed that, in order to obtain realistic positive curvature, the following characteristics should be present in a chosen function:
\begin{enumerate}
    \item The function should be peaked at $\theta = 0$;
    \item The peak should be wide enough so that not all the pressure is focusing at $\theta = 0$, which would give negative curvature;
    \item While the peak needs to be wide, the function cannot fully wrap the droplet to $\theta = \pi$, which would stop expansion. This requires the function to have a wide peak, but short tail. 
\end{enumerate}
The second condition can be easily achieved with the classical Gaussian profile by increasing the $\sigma$ parameter, which widens the peak. However, Gaussian profiles are known for having a very large tail, such that widening the peak would automatically break the third condition, as the profile would wrap the whole droplet. In statistics, the ``tail size" of a distribution is often characterized by excess kurtosis. A Gaussian function has an excess kurtosis of exactly 0 and is referred to as a mesokurtic function. Leptokurtic functions are those with a positive excess kurtosis, and these are not desired in this study since they lead to fat-tailed profiles. Platokurtic functions have negative excess kurtosis, which leads to a short tail. Therefore, platokurtic functions are optimal for our study in order to guarantee the third condition above. Out of many traditional statistical distributions, the raised cosine is one of the few smooth functions that provide a constant and negative excess kurtosis ($Kurt = -0.59$) that does not depend on the parameter $W$. Due to its simplicity, while also being a classical function that provides the necessary kurtosis condition, we chose to use the raised cosine in this study. Throughout this work we will show that the raised cosine can, indeed, be a suitable choice to simulate situations in which the deformed sheet presents a positive curvature.

\begin{table}[htbp]
\centering
\begin{tabular}{ |c|c|c| }
    \hline
    Profile name & Expression & Control parameter \\
    \hline
    & & \\ [-5pt]
    Cosine & $\cos\left({\theta}\right) H\left(\frac{\pi}{2} - \theta\right)$ & - \\ 
    & & \\ [-5pt]
    Gaussian & $\exp{\left(\frac{x^2}{2\sigma^2}\right)}$  & $\sigma$ \\ 
    & & \\ [-5pt]
    Raised cosine & $\frac{1}{2}\left( 1 + \cos\left(\theta \frac{\pi}{W}\right) \right) H(W - \theta)$ & $W$ \\ [5pt]
    \hline
\end{tabular}
\caption{Different initial pressure profiles used throughout this work. In the cosine and raised cosine expressions, $H$ represents the standard Heaviside step function. }
\label{table:pressure_profiles}
\end{table}

This paper is organized as follows. Section \ref{sec:experimt_setup} introduces the experimental setup used. In Section \ref{sec:numerical_approach}, the problem is mathematically stated, and the numerical approach to simulate the droplet deformation is described.  Section \ref{sec:all_results} discusses the numerical and experimental results obtained. Finally, section \ref{sec:conclusion} concludes the results and presents future perspectives. Additional numerical details can be found in the appendices.

\section{Experiment}
\label{sec:experimt_setup}
Our experimental setup has previously been described in detail in \cite{kurilovich_plasma_2016,kurilovich_laser-induced_2019,liu_laser-induced_2021}. 
In short, in these experiments, a \SI{}{\kHz} sequence of liquid tin droplets (\SI{270}{\degreeCelsius}) are vertically dispensed in a vacuum environment (\SI{e-7}{mbar}) by a droplet generator. The pressure and frequency of the generator are adjusted to obtain droplets with diameters ranging from \SI{27}{\um} to \SI{59}{\um}. The droplets fall at approximately \SI{10}{\meter / \second} through a horizontal light sheet generated by a continuous-wave HeNe laser. The scattered light is then detected by a photomultiplier tube and downsampled to \SI{10}{\Hz}, serving as the start trigger for the experiment. 

Firstly, a droplet is hit by the Gaussian intensity pre-pulse (PP) [$\lambda =$ \SI{1064}{\nano\meter}, FWHM = \SI{10}{\ns}, circularly polarized]. This laser interacts with the droplet by ablating part of its surface and generating a plasma, exerting pressure on the remaining liquid tin. This pressure rapidly propels and expands the tin droplet on the order of \SI{100}{\meter / \second} to a thin axisymmetric sheet \cite{kurilovich_plasma_2016,kurilovich_power-law_2018}.
The propulsion velocity $U_z$ aligns with the propagation direction of the laser, while the expansion occurs radially with an initial velocity $\dot{R}_{0}$. The radial expansion speed subsequently decreases due to the surface tension that is exerted on the edge of the sheet \cite{villermaux_drop_2011, Gelderblom2016, kurilovich_plasma_2016}.

\begin{figure}[H]
\centering
\includegraphics[width=0.9\textwidth]{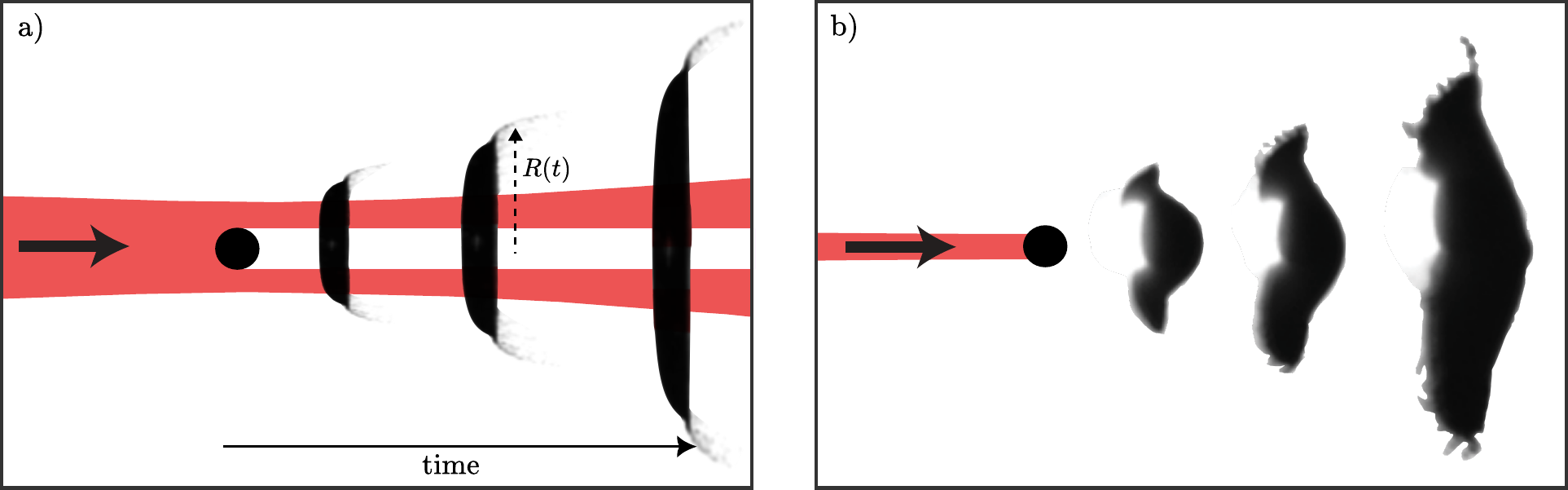}
\caption{Laser pulse schemes with specific irradiation geometries lead to either a) positive or b) negative curvature, visualized by shadowgraphs. }
\label{fig:sketch1_experimental_setup}
\end{figure}

The timescales governing propulsion and expansion accelerations are similar to the duration of the laser pulse (\SI{}{\ns}) and are much shorter than the timescale of the subsequent fluid dynamics deformation \cite{Gelderblom2016, kurilovich_plasma_2016}.
Figure \ref{fig:sketch1_experimental_setup}a illustrates the typical experimental response of droplets to the pre-pulse (PP) impact, characterized by a beam width of approximately \SI{100}{\micro\meter} FWHM and showcasing a sheet that curves away from the laser beam. 
Conversely, figure \ref{fig:sketch1_experimental_setup}b depicts the response to a PP with a beam diameter of approximately \SI{20}{\micro\meter}, exhibiting negative curvature.

In order to observe the evolution of the liquid tin, we obtain shadowgraphs as depicted in figure \ref{fig:sketch1_experimental_setup}. The shadowgraphy imaging setup is described in detail in \cite{kurilovich_laser-induced_2019} and briefly summarized here. It consists of a dye-based illumination source and a charge-coupled camera coupled to a long-distance microscope, granting a spatial resolution of approximately \SI{5}{\um}.
The illumination source emits pulses with a duration of \SI{5}{\ns} (FWHM) and a spectral bandwidth of \SI{12}{\nm} (FWHM) centered at \SI{560}{\nm}. We utilize these shadowgraphy pulses (SP) for backlight illumination of the side-view acquisitions (at \SI{90}{\degree} concerning the laser axis) to capture the curvature dynamics of the liquid tin sheet.
We select a discrete number of time steps in the experiment and record 20 frames in a stroboscopic manner for each step, each frame representing a distinct laser-droplet interaction event. This methodology allows us to apply post-filtering techniques, such as selecting optimally aligned laser-to-droplet events.

\section{Problem formulation and numerical approach}
\label{sec:numerical_approach}
We will perform fluid simulations to numerically simulate the laser-induced deformation of a tin droplet into a sheet. The governing equations for the isothermal incompressible bi-phase (droplet and air) flow are the continuity and momentum conservation, given by
\begin{align}
& \rho\left( \frac{\partial \Ubold}{\partial t} + \nabla \cdot ( \Ubold\Ubold ) \right)	= - \nabla p  + \mu \nabla^2\Ubold  + \mathbf{f}_{\gamma}, \label{eq:navier1} \\
& \nabla \cdot \Ubold = 0, \label{eq:navier2}
\end{align}
where $\Ubold$ and $p$ are the velocity and pressure fields, and $\rho$ and $\mu$ are the fluid density and viscosity, respectively. In the numerical method used here, the surface tension force is defined as a body force $\mathbf{f}_{\gamma} = \gamma \kappa \delta_s \textbf{n}$, where $\kappa$ is the local curvature of the interface, $\gamma$ the constant surface tension coefficient, $\textbf{n}$ is the unit vector normal to the interface, and $\delta_s$ is the Dirac delta function centered on the interface~\cite{popinet2009accurate, Tryggvason2011-book}.

Equations \eqref{eq:navier1}-\eqref{eq:navier2} can be nondimensionalized by rescaling variables with the following choices:

\begin{equation}
    \textbf{x} = R_0\bar{\textbf{x}}, \hspace{10pt} t = \frac{R_0}{U_z}\bar{t}, \hspace{10pt} \Ubold = U_z\bar{\Ubold}, \hspace{10pt} p =  \rho \,U_z^2\bar{p}, \hspace{10pt} \kappa = \frac{1}{R_0}\bar{\kappa}, \hspace{10pt} \delta_s = \frac{1}{R_0}\bar{\delta_s},
\label{eq:nondim_scales}
\end{equation}
where $R_0$ is the initial radius of the droplet, and $U_z$ is the droplet propulsion velocity obtained after the laser hit.

Substituting \eqref{eq:nondim_scales} into \eqref{eq:navier1}-\eqref{eq:navier2}, we obtain the non-dimensional version of the governing equations, given by
\begin{align}
& \frac{\partial \bar{\Ubold}}{\partial t} + \nabla \cdot ( \bar{\Ubold}\bar{\Ubold} ) 	= - \nabla \bar{p}  + \frac{1}{Re} \nabla^2\bar{\Ubold}  + \frac{1}{We} \bar{\kappa} \bar{\delta_s} \textbf{n}, \label{eq:nondim_navier1} \\
& \nabla \cdot \bar{\Ubold} = 0, \label{eq:nondim_navier2}
\end{align}
where 

\begin{equation}
   Re = \frac{\rho \, U_z \, R_0}{\mu}, \quad \mathrm{and} \quad We = \frac{\rho \, U_z^2 \, R_0}{\gamma}
\label{eq:nondim_numbers}
\end{equation}

are the Reynolds and Weber numbers, respectively. Later in the text we also use a version of these two numbers based on the droplet initial expansion velocity $\dot{R}_0$ as velocity scale, that is, $Re_{def} = \frac{\rho \dot{R}_0 R_0}{\mu}$ and $We_{def} = \frac{\rho  \dot{R}_0^2 R_0}{\gamma}$. 

Equations \eqref{eq:nondim_navier1}-\eqref{eq:nondim_navier2} require an appropriate velocity field as the initial condition to be solved. To obtain an initial condition for the velocity, we adopt the same strategy proposed by \cite{Gelderblom2016}, which we briefly describe here. We know that the process is nearly inviscid and that the timescale of this plasma kick $(\tau_p)$ is much smaller than the inertial $(\tau_i)$ and capillary $(\tau_c)$ timescales, that is $\tau_{p} \ll \tau_{i} \ll \tau_{c}$. Therefore, we assume that the tin plasma generated in the process applies a near-instantaneous pressure kick on the droplet surface. This pressure kick will be provided as a function $f(\theta)$ over the parametrized surface of the droplet as in figure \ref{fig:initialization_strategy}.  With these assumptions, the Navier-Stokes equations can be reduced into the following Laplace equation for the pressure
\begin{equation}
\nabla^2 \bar{p} = 0.
\label{eq:laplace_pressure}
\end{equation}

Equation \eqref{eq:laplace_pressure} is solved in spherical coordinates assuming symmetry in the azimuthal angle $\varphi$. Therefore, it is solved in the domain $(r, \theta) \in [0, 1] \times [0, \pi]$ with boundary condition $\bar{p}(1, \theta) = f(\theta)$. A solution to \eqref{eq:laplace_pressure} can be obtained by decomposing the pressure field in Legendre polynomials $P_n$, which will give
\begin{equation}
    \bar{p}(r, \theta) = \sum_{n=0}^{\infty} A_n r^n P_n( \cos\theta ),
\label{eq:legendre_pressure}
\end{equation}
with coefficients
\begin{equation}
    A_n = \frac{2n + 1}{2}\int_0^\pi f(\theta) P_n( \cos\theta ) \sin\theta.
\label{eq:legendre_pressure}
\end{equation}

The initial velocity field within the droplet can then also be obtained from the simplified Navier-Stokes equations by
\begin{equation}
\bar{\textbf{u}}_{0} = - \alpha \nabla \bar{p} = -\alpha\left[ n \sum_{n=0}^{\infty} A_n r^{n-1} P_n( \cos\theta ), \ \ - \sum_{n=0}^{\infty} A_n r^n P'_n( \cos\theta ) \sin(\theta) \right],
\label{eq:init_velocity}
\end{equation}
where $\alpha$ can be numerically tuned to change the magnitude of the velocity field. Since we opted to use the propulsion speed as our velocity scale in \eqref{eq:nondim_scales}, we will always take $\alpha$ such that the nondimensional propulsion velocity is 
\begin{equation}
    \bar{U}_z = \frac{3}{2} \int_0^\pi f(\theta) \cos\theta \sin\theta = 1.
\label{eq:propulsion_velocity}
\end{equation}
The velocity field $\bar{\textbf{u}}_{0}$ from \eqref{eq:init_velocity} is then used as an initial condition for equations \eqref{eq:nondim_navier1}-\eqref{eq:nondim_navier2}.

Equations \eqref{eq:nondim_navier1}-\eqref{eq:nondim_navier2} are solved numerically using the open-source free-software language Basilisk C \cite{Popinet2013-Basilisk}. The droplet is created at the center of a square domain $[-10R_0, \ 10R_0] \times [-10R_0, \ 10R_0]$ which is fully discretized with a non-uniform quadtree grid~\cite{popinet2003gerris,popinet2009accurate}. To accurately resolve the flow structure inside the droplet and its shape, we apply increased refinement levels for the liquid phase and also at the interface. The maximum quadtree level of refinement used is 14, resulting in grid cells with a minimum size of $\Delta = 20R_0/(2^{14}) = 0.0012R_0$. 

The interface between the droplet and the outside fluid is represented implicitly by the Volume of Fluid (VOF) scheme \cite{Hirt1981}, in which each mesh cell stores a value representing the fraction of droplet fluid. Density and viscosity are then locally determined based on the volume fraction $c(\textbf{x}, t)$ according to
\begin{align}
    \rho(c) = & \  c \ \rho + (1 - c)\rho_a, \\
    \mu(c) = & \ c \  \mu + (1 - c)\mu_a.
\label{eq:density_viscosity}
\end{align}
where $\rho_a$ and $\mu_a$ represent the density and dynamic viscosity of the outer fluid, respectively. While in experiments the droplet is contained within a quasi-vacuum chamber, due to numerical limitations, we keep the outer fluid properties set to $\rho_a = 10^{-4} \rho$ and $\mu_a = 10^{-4} \mu$. The volume fraction field $c$ is then advected over time by solving the equation
\begin{equation}
    \del{c}{t} + \nabla \cdot (c \textbf{u}) = 0.
\label{eq:vof_advection}
\end{equation}

The numerical code then solves the governing equations using a projection method and a multilevel Poisson solver. We refer to \cite{popinet2009accurate,Popinet2015} for more details of the VOF implementation and {Appendix~\ref{section:appendix_validation} for validation tests of our code}. We note that the software language Basilisk C used here has also been previously validated in many other works involving deformable surfaces, such as \cite{Sanjay2021, popinet2009accurate}. The full code used to perform the simulations in this work can be accessed in [LINK TO BE ADDED DURING PUBLICATION].

\section{Results and discussion}
\label{sec:all_results}

\subsection{Predicting curvature based on the initial velocity field}
\label{sec:predicting_curvature}

We begin our study by looking at the initial velocity calculated at time $t = 0$ from equation \eqref{eq:init_velocity} and attempting to predict how the sheet will curve after deformation.

In order to do this, we first subtract the center-of-mass velocity from $\bar{\Ubold}_{0}$ such that we obtain only the velocity field responsible for the droplet deformation. From \eqref{eq:init_velocity} and \eqref{eq:propulsion_velocity}, this field is given by

\begin{equation}
\bar{\textbf{u}}_{def}(r, \theta) = \bar{U}_z\left[\cos\theta, \ \sin\theta \right]  + \left[ - n \sum_{n=0}^{\infty} A_n r^{n-1} P_n( \cos\theta ), \ \sum_{n=0}^{\infty} A_n r^n P'_n( \cos\theta ) \sin(\theta) \right].
\label{eq:deformation_velocity}
\end{equation}

An indicative of how the droplet will deform is the radial component of $\bar{\textbf{u}}_{def}$ at the droplet surface, that is
\begin{equation}
\bar{u}_{r_{def}}(1, \theta) = \bar{U}_z\cos\theta  - n \sum_{n=0}^{\infty} A_nP_n( \cos\theta ).
\label{eq:surface_deformation_velocity}
\end{equation}

\begin{figure}[htbp]
\centering
\includegraphics[width=1\textwidth]{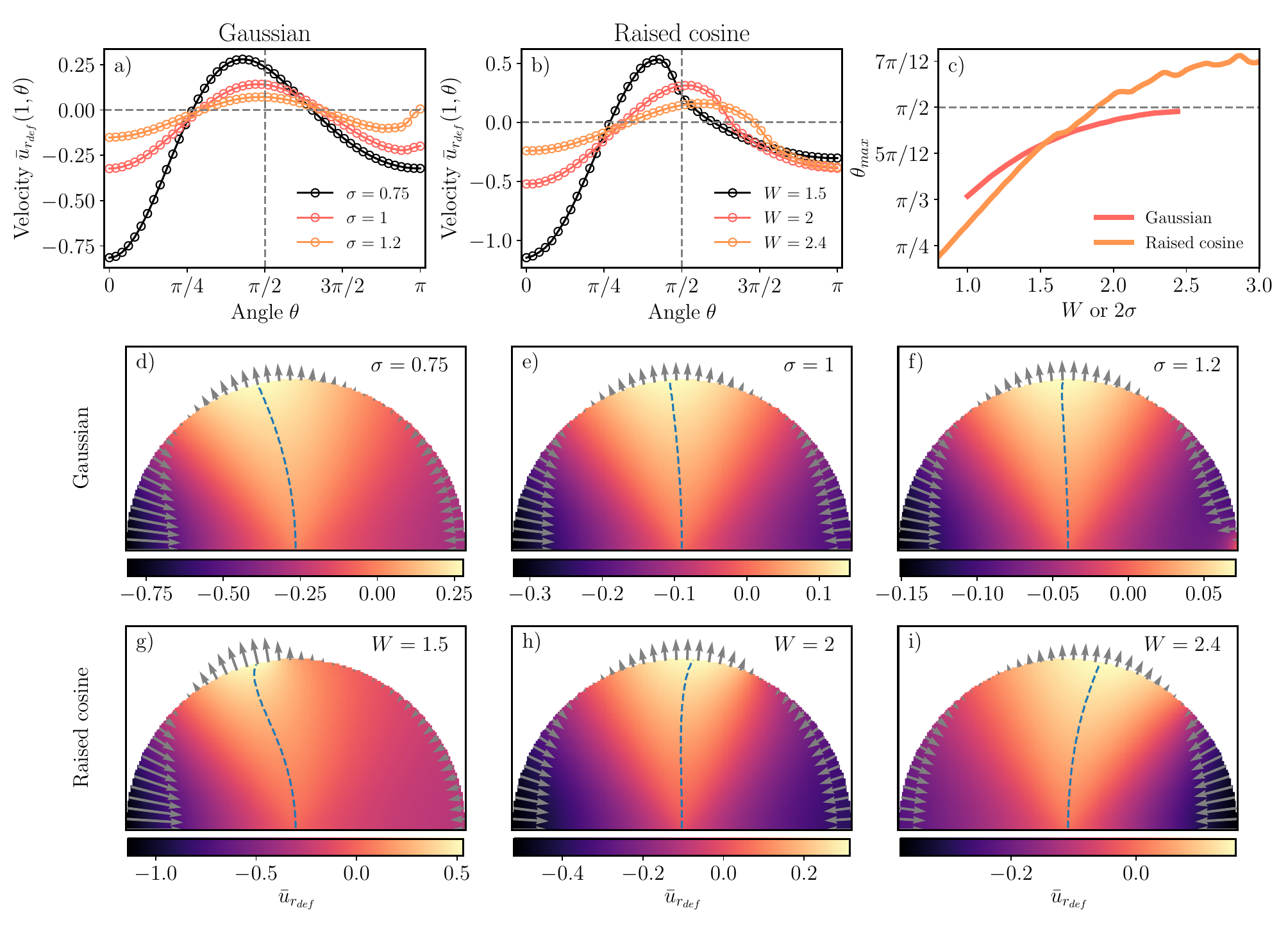}
\caption{Initial droplet radial velocity $\bar{u}_{r_{def}}(r, \theta)$ in spherical coordinates at time $t = 0$ for different profiles and profile parameters. Panels a-b) velocity at the droplet surface over $\theta$ for three Gaussian and three Raised Cosine profiles, respectively. Panel c) Value of $\theta_{max}$ as a function of profile parameter. Panels d-i) $\bar{u}_{r_{def}}$ field within the droplet for the 6 cases shown in panels a-b. The arrows represent the magnitude and sign of $\bar{u}_{r_{def}}(1, \theta)$ and the dashed line indicates the point of maximum $\bar{u}_{r_{def}}$ for each $r$.}
\label{fig:ur_over_theta}
\end{figure}

Figures \ref{fig:ur_over_theta}a-b show the value of the radial velocity \eqref{eq:surface_deformation_velocity} as a function of $\theta$ for three Gaussian and three raised cosine profiles. Intuitively, we see negative velocities for low and high $\theta$, with positive values at intermediate $\theta$. This is expected since the droplet is being ``squeezed" in the laser-hit direction while it expands upwards. A striking feature of the Gaussian curves is that the high values of $\bar{u}_{r_{def}}$ always happen before $\pi / 2$, regardless of the choice for the parameter $\sigma$. This is also shown in figures \ref{fig:ur_over_theta}d-f, where the arrows represent $\bar{u}_{r_{def}}$ on the droplet surface. This observation indicates that the maximum radial expansion for a Gaussian profile will always happen behind the droplet center of mass, such that a negative curvature is to be expected. We note that this had also previously been observed in \cite{Gelderblom2016}, where the authors show a convergence of the direction of maximum expansion to an angle close to $\pi/2$.

For a raised cosine profile, on the other hand, we can see that the maximum value of $\bar{u}_{r_{def}}$ can be positioned beyond $\pi/2$ if $W$ is chosen to be large enough. This can be seen in figures \ref{fig:ur_over_theta}b, g-i. Since the maximum radial expansion happens after the droplet center-of-mass, we expect the deformed sheet to have a positive curvature. It is worth noticing that the negative curvature can also be expected from raised cosine profiles, as long as $W$ is chosen small enough. This highlights the versatility of this pressure profile choice, as both outcomes can be achieved with the correct tuning of a single parameter.

To quantitatively predict if a profile can provide positive or negative curvature, we can calculate the angle where $\bar{u}_{r_{def}}(1, \theta)$ is maximized, let us call that angle $\theta_{max}$. We can find this local maximum by derivating \eqref{eq:surface_deformation_velocity} and solving the equation
\begin{equation}
\bar{u}'_{r_{def}}(1, \theta) = - U_z\sin\theta  + \sum_{n=0}^{\infty} n \sin\theta A_n P_n'( \cos\theta ) = 0.
\label{eq:deriv_surface_velocity}
\end{equation}
The solution to this equation is shown in figure \ref{fig:ur_over_theta}c for the Gaussian and raised cosine profiles as a function of their respective parameters, $\sigma$ and $W$. Once again, we see more clearly that the Gaussian solution cannot cross the $\theta_{max} = \pi/2$ limit, while the raised cosine does. We note that the Gaussian results are only shown up to $\sigma \approx 1.2$, since for higher values, the Gaussian pressure profile wraps the entire droplet such that no significant expansion happens.

We can also show that a Gaussian pressure profile can never provide $\theta_{max} > \pi/2$. If we calculate the derivative \eqref{eq:deriv_surface_velocity} at the point $\theta = \pi/2$, we have
\begin{equation}
\bar{u}'_{r_{def}}\left(1, \frac{\pi}{2}\right) = - U_z  + \sum_{n=0}^{\infty} n A_n P_n'(0).
\label{eq:deriv_halfpi_surface_velocity}
\end{equation}
If the value of \eqref{eq:deriv_halfpi_surface_velocity} is positive, we know the function is still increasing at $\pi/2$, such that the maximum velocity will happen at $\theta_{max}>\pi/2$. Analogously, if the value is negative, we will have $\theta_{max}<\pi/2$. If we substitute the Gaussian profile into $A_n$ and $\bar{U}_z$ in \eqref{eq:deriv_halfpi_surface_velocity}, we obtain
\begin{equation}
\bar{u}'_{r_{def}}\left(1, \frac{\pi}{2}\right) = - \frac{3}{2}\int_0^\pi \cos\theta \sin\theta e^{
-\theta^2/(2\sigma^2) }\text{d}\theta + \sum_{n=0}^{\infty} n \frac{2n + 1}{2} P_n'(0) \int_0^\pi P_n( \cos\theta ) \sin\theta e^{
-\theta^2/(2\sigma^2) } \text{d}\theta,
\label{eq:halfpi_surface_velocity_gaussian}
\end{equation}
which is always negative.

The radial deformation velocity from equation \eqref{eq:surface_deformation_velocity} can also be defined for other fixed radial positions (instead of only at the droplet surface). By doing that for multiple values of $r$ and storing the angle at which the maximum velocity is attained, we obtain the dashed lines shown in figure \ref{fig:ur_over_theta}d-i. This line indicates the direction of maximum radial expansion not only at the droplet surface but also within the fluid. This shows that the expansion direction smoothly varies from the droplet center to the surface. An almost-flat curve is obtained for a Gaussian with $\sigma = 1.2$ and $W\approx 2$, indicating that these droplets will expand almost perfectly vertically for all $\theta$ cross sections.

\subsection{Curvature measurements after droplet deformation}
The parameter $\theta_{max}$ was introduced as an initial condition indicator of the sheet curvature characteristics in the previous section. In this section, we will attempt to demonstrate that this parameter can indeed be correlated with the actual later-time sheet curvature. Since we have demonstrated in section \ref{sec:predicting_curvature} that Gaussian profiles can never provide positive curvature, we will then focus our study now only on raised cosine profiles.

Multiple simulations were performed varying the width parameter in the range $W \in [1.0, 3.0]$. Lower values of $W$ result in numerical difficulties in the convergence of $\eqref{eq:init_velocity}$, while for higher values, no significant expansion can be observed, as the pressure fully wraps the droplet as also noticed by \cite{Gelderblom2016} for wide Gaussian profiles. The deformation Reynolds and Weber numbers were always kept high enough ($Re_{def}>1000$ and $We_{def} > 1000$), such that no significant viscous effects are present, and significant capillary effects are only observed at later times in the simulations (particularly the formation of a rim at the edge of the sheet). None of these effects are expected to affect the bulk curvature formation of the sheet.

\begin{figure}[htbp]
\centering
\includegraphics[width=1.0\textwidth, trim={0 60 0 45},clip]{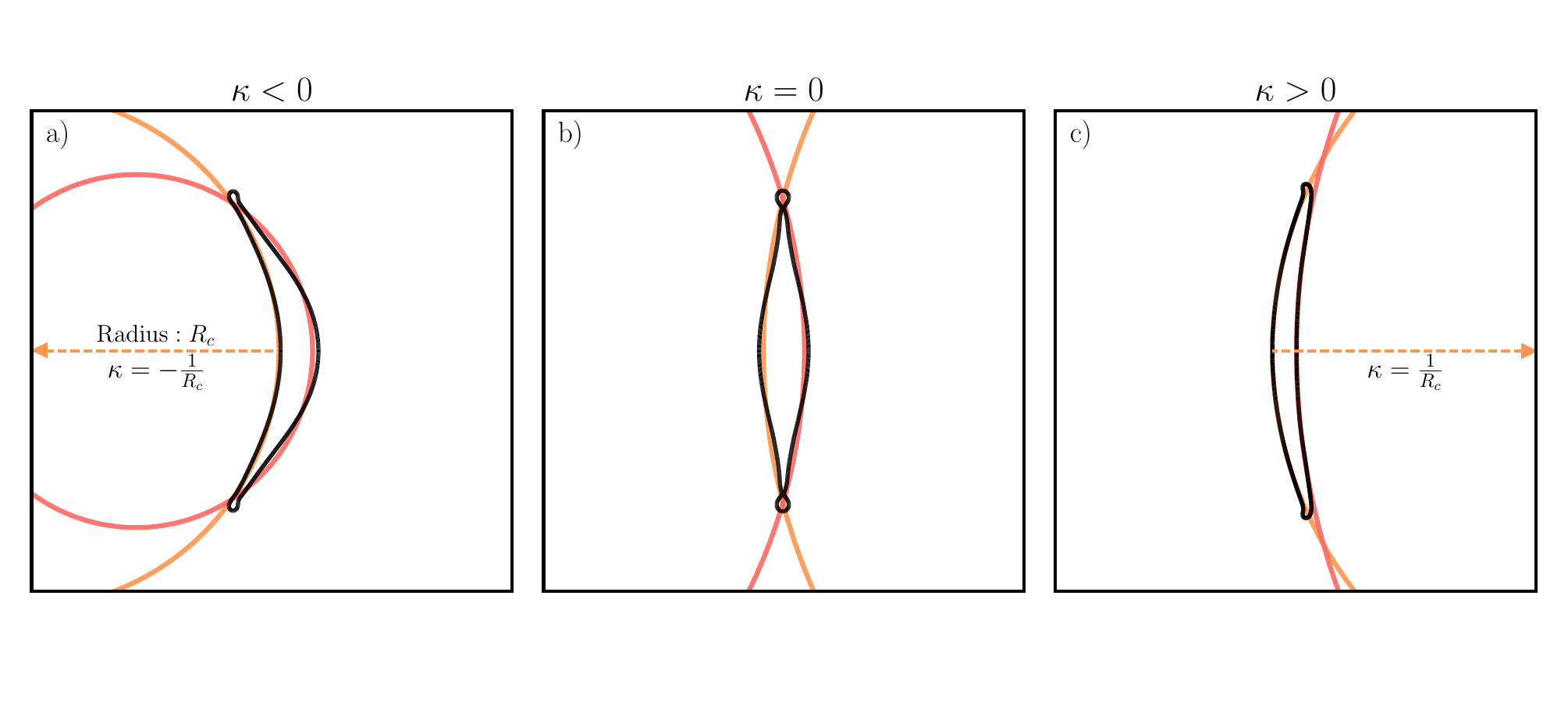}
\caption{Method used to define and extract the sheet curvature of experiments and simulations based on fitted circles. The black lines represent the sheet interface, as extracted from a simulation or an experiment. The orange and red lines show the circles that were fitted to the laser-side and back-side of the sheet, respectively.}
\label{fig:curv_circle_fit_method}
\end{figure}

To quantify the curvature of the deformed sheet, we extract the sheet cross-section from the simulations and perform circular fits. This process is illustrated in figure \ref{fig:curv_circle_fit_method}. The two sides of the sheet are separately extracted, and individual circles are fitted to each side, resulting in two curvatures $\kappa_{1}$ and $\kappa_{2}$. Each value will be defined as negative if the corresponding circle center is in the same direction as the laser or positive otherwise. We then define the simulated sheet curvature as
\begin{equation}
    \kappa_{avg} = \frac{\kappa_{1} + \kappa_{2}}{2}.
\label{eq:definition_kappa_simulation}
\end{equation}
With this definition, we will obtain $\kappa = 0$ if the sheet is perfectly symmetric, as in the case of the cosine profile shown previously in figure \ref{fig:initialization_strategy}. It will be positive when the sheet curves away from the laser origin and negative when it curves towards the laser origin. These three possibilities are illustrated in figure \ref{fig:curv_circle_fit_method}, showing the fitted circles used to estimate the curvatures.

\begin{figure}[htbp]
\centering
\includegraphics[width=1\textwidth]{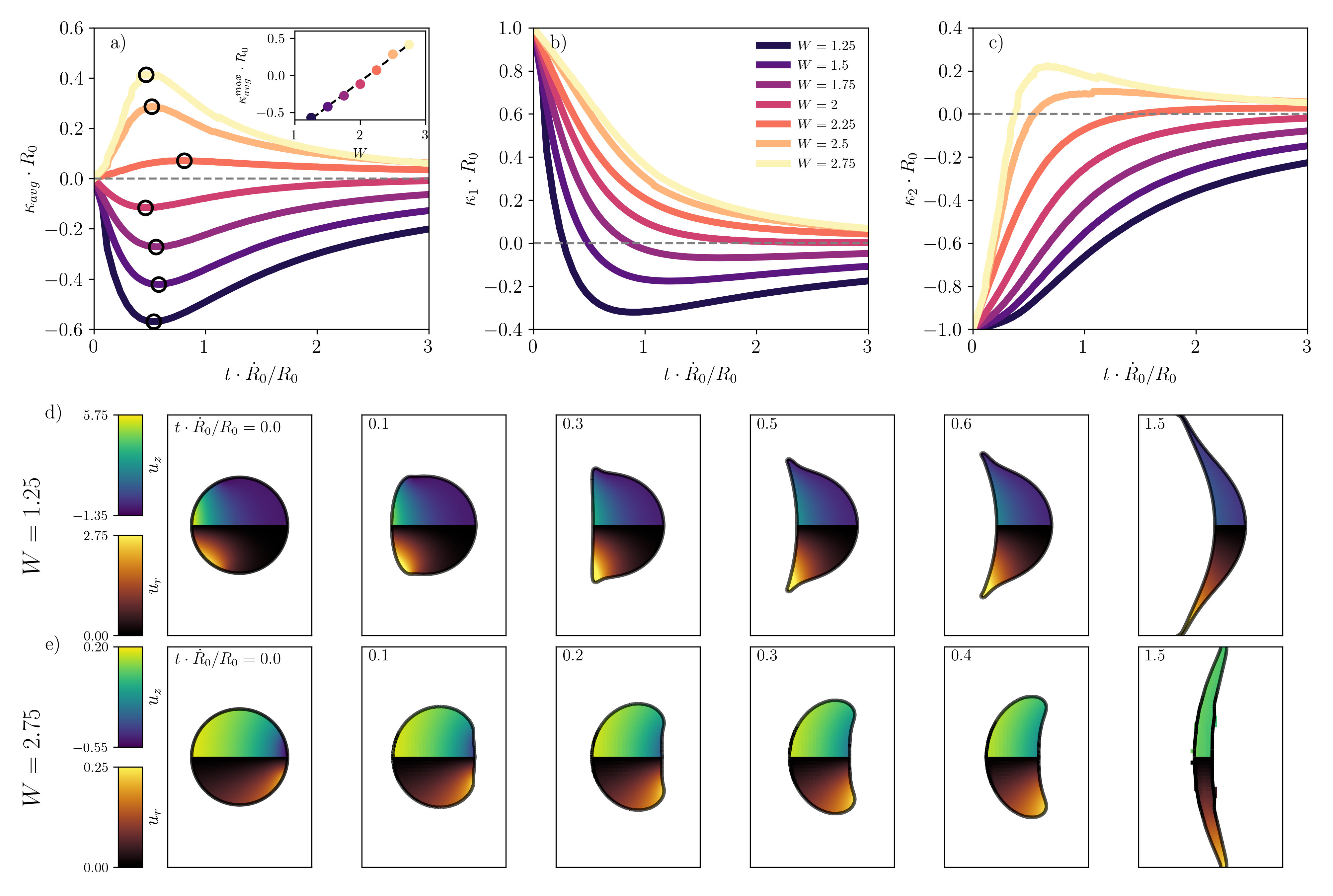}
\caption{Sheet curvature over time for different values of the raised cosine parameter $W$. Panel a) average curvature $\kappa_{avg}$, b) $\kappa_1$ and c) $\kappa_2$. The inset shows the peak average curvature $\kappa^{max}_{avg}$ for each simulation as a function of $W$. Panels d-e) droplet/sheet interface at selected timestamps for two simulations with $W = 1.25$ and $W = 2.75$, respectively. The components of the velocity field in cylindrical coordinates $u_r$ and $u_z$ are also shown in the snapshots.}
\label{fig:transient_curvature}
\end{figure}

We begin by looking at how the numerical curvatures $\kappa_1$, $\kappa_2$ and $\kappa_{avg}$ develop over time. Figure \ref{fig:transient_curvature}a presents $\kappa_{avg}$ over nondimensional time for different raised cosine profiles. The full evolution of all simulations in this figure can also be seen in video 1 of the supplementary material. Initially, all curves start from the origin since $\kappa_{avg} = 0$ in the case of a perfect sphere. Over time, each simulation reaches a peak curvature as the droplet experiences its initial deformation, which then tends again to zero as the sheet expands and becomes flatter. While some curves show positive values and others negative, none present a change of sign over time. This means that $\kappa_{avg}$ is a good theoretical quantity to classify if a simulation results in positive or negative curvature since this classification will not depend on the chosen measurement time.

For low values of $W$, as illustrated in figure \ref{fig:transient_curvature}d, the focused profile quickly pierces the laser-side of the droplet, changing it from its initial convex shape to a concave shape, which results in negative values for $\kappa_{avg}$. Due to the focused profile, we can see that most of the velocity field is concentrated at the laser-side of the droplet, which deforms very fast and creates the typical negative curvature shape. This initial curvature formation happens in approximately half the expansion (inertial) timescale, as can be seen in figure \ref{fig:transient_curvature}a by the moment when a peak in $\kappa_{avg}$ is obtained. After this peak is reached, most of the deformation observed will be simple sheet expansion, such that the curvature will again lower in magnitude as the sheet becomes more extended. This initial curvature deformation is seen very clearly by looking at the initial field $u_z$, also shown in figure \ref{fig:transient_curvature}d for $W = 1.25$. We observe that $u_z$ has only a small concentrated area of strong positive values at the laser-side of the droplet. The opposite side of the droplet maintains mostly its spherical shape during this stage. Over time, due to mass conservation, the $u_z$ component is converted into $u_r$ as the droplet is progressively ``squeezed" in the $z$-direction and expands in $r$. This causes the $u_r$ field to become closer to uniform within the droplet, as seen in figure \ref{fig:transient_curvature}d. By observing the velocity component field $u_r$ for $W =1.25$ we can also see a good indication that negative curvature will be obtained, since the higher values of $u_r$ are strongly concentrated on the laser side of the droplet at initial time. Due to this, the droplet expands towards the laser, and the curvature becomes increasingly negative until it eventually reaches its peak.

On the other hand, the unfocused pressure profile wraps the droplet so that the laser-side of its surface does not experience strong deformation and maintains its approximate initial spherical shape. This can also be seen clearly through the velocity field $u_z$ in figure \ref{fig:transient_curvature}e for $W = 2.75$. Most of the early deformation happens on the opposite side of the droplet, where a strong negative velocity field pushes the surface inwards, creating a positive curvature. The velocity field $u_r$ also indicates positive curvature, since the highest values of $u_r$ are concentrated on the droplet side not illuminated by the laser. Similarly to the previous case, we see also that the curvature experiences a peak. Once again, this is set by the time it takes for the initial velocity field to establish the shape of the sheet, after which only expansion with a fixed shape will be present. 

The inset in figure \ref{fig:transient_curvature}a presents the value of this peak ($\kappa^{max}_{avg}$) as a function of the profile parameter $W$. Interestingly, we see a linear growth of the peak curvature within this range of $W$. Another interesting observation is that the timescale of the peak formation is not heavily dependent on $W$ and approximately half the inertial timescale. Due to the similar timescale magnitude, the processes of sheet curving and sheet expansion happen simultaneously and cannot be easily separated over time.

Figures \ref{fig:transient_curvature}b-c also show the individual curvatures $\kappa_1$ and $\kappa_2$ for the laser-side and opposite-side of the droplet, respectively. Unlike $\kappa_{avg}$, these individual curvatures can change sign over time. The curvature $\kappa_1$ will always present a sign change for simulations with small $W$. This happens as initially $\kappa_1 = 1$ due to the spherical shape and eventually becomes negative as the laser-side of the droplet is strongly pushed by a focused pulse. However, $\kappa_2$ will not change sign for simulations of small $W$ since the opposite side of the droplet does not experience strong pressure to deform, and its initial curvature does not change direction. On the other hand, for simulations with high $W$ the exact opposite happens, that is, $\kappa_2$ changes sign over time while $\kappa_1$ does not. 

As suggested by figure \ref{fig:transient_curvature}, a direct correlation between $\kappa_{avg}$ and $W$ will vary with time, however the sign change will not be time-dependent. Since our main interest is to know when a sheet will be positively or negatively curved, we will now study this correlation between $\kappa_{avg}$ and $W$ at a fixed time.

\begin{figure}[htbp]
\centering
\includegraphics[width=\textwidth]{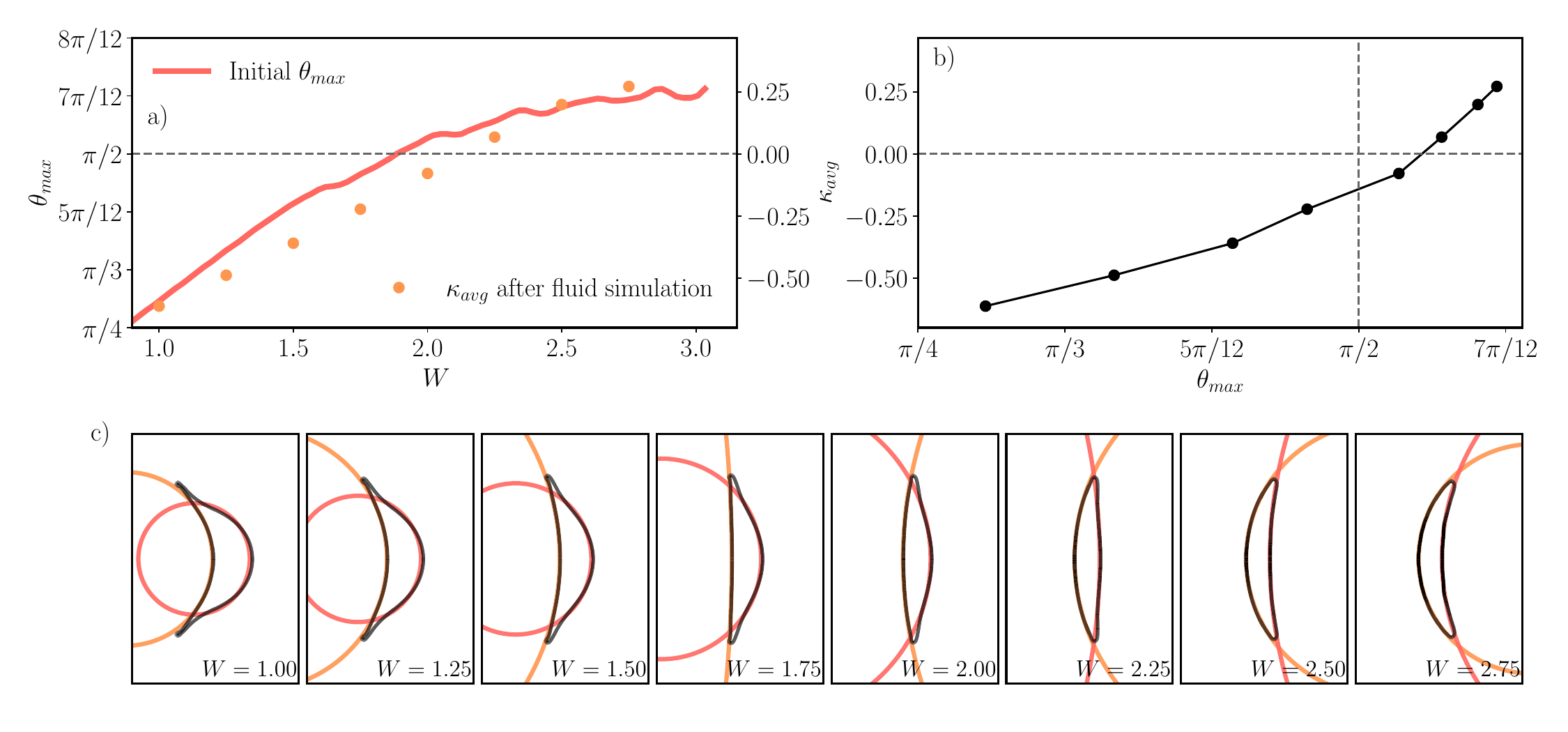}
\caption{Simulated average curvature and $\theta_{max}$ for different raised cosine parameters $W$. The curvature is always obtained at time $t \cdot \dot{R}_0/R_0 = 1$. Panel a) $\theta_{max}$ and $\kappa_{avg}$ over $W$. Panel b) correlation between $\theta_{max}$ and $\kappa_{avg}$. Panel c) snapshots for data points shown in panels a-b. The colored lines represent the two circle fits in figure \ref{fig:curv_circle_fit_method}.}
\label{fig:initial_deformed_curv_correlation}
\end{figure}

In figure \ref{fig:initial_deformed_curv_correlation} we present the measured value of $\kappa_{avg}$ as a function of the profile width $W$ at a specific time of the simulations $t \cdot \dot{R}_0/R_0 = 1$, where $\dot{R}_0$ is the initial expansion velocity experienced by the sheet. The timescale $\dot{R}_0/R_0$ is then an inertial timescale related to how fast the sheet expands. We chose to match this specific nondimensional time for all simulations so that all sheets have the same amount of expansion at the moment of measurement. As expected, we see that the curvature starts at negative values for focused pressure profiles (low $W$) and then monotonically increases with $W$, eventually approaching a plateau. In the same figure, we visualize the values of $\theta_{max}$ introduced in figure \ref{fig:ur_over_theta}. We can observe that both $\theta_{max}$ and $\kappa_{avg}$ follow qualitatively a very similar trend. Moreover, the $\theta_{max}$ curves indicate that the curvature will flip at $W = 1.9$, and the $\kappa_{avg}$ curve indicates the flip around $W = 2$, which is a reasonable agreement. 

To visualize the correlation between the predicted and measured curvatures, we plot in figure \ref{fig:initial_deformed_curv_correlation}b the value $\theta_{max}$ versus $\kappa_{avg}$. We observe that the curve passes not too far from the point $(\theta_{max}, \kappa) = (\pi/2, \ 0)$, which is the point that indicates a perfect prediction of the curvature flip. For illustration, in the bottom row of figure \ref{fig:initial_deformed_curv_correlation}, we show the sheet cross-section for eight of the simulations used in the other panels. One can easily see from these snapshots that the curvature sign flips around $W = 2$. Other interesting characteristics can also be seen in these snapshots outside the curvature measurements. We observe, for example, that the thickness profile along the sheet is closer to constant for high values of $W$, while the sheet is very thick at its center and thin at the edges for low $W$. These thin edges will eventually numerically disconnect from the sheet, forming rings, as our grid is not fine enough to resolve it properly. We note that, in experiments, we do indeed observe small droplets fragmenting near the edge of the sheet. However, this is not what is observed here since our simulation assumes axisymmetry and asymmetric fragments cannot be reproduced. The near-constant thickness profile obtained from high $W$ is beneficial for the experimental application of EUV-light generation since the tin over the whole sheet can be evenly ablated into plasma by a following laser pulse.

\subsection{Comparison with experiments}
We have seen so far that the raised cosine profile can provide deformed sheets with a curvature that goes from negative to positive as $W$ is changed. We will now attempt to visualize how well this can be correlated to the sheet shapes obtained experimentally when the ratio between laser focus and droplet size is changed.

To quantify the sheet curvature experimentally, we will use the side-view shadowgraphs obtained following the experimental method described in section \ref{sec:experimt_setup}. For a given shadowgraph, we apply an edge-extraction algorithm to obtain the interface of the sheet. We then perform two circle fits as described previously in figure \ref{fig:curv_circle_fit_method} for simulations. However, unlike in the simulations, we note that the shadowgraphs do not show a cross-section of the sheet. They only show a projection of the three-dimensional sheet into the camera plane. Consequently, if the sheet has a curvature, the concave side of the interface will look flat, and the actual curvature of that side will be hidden in the projection. With this limitation in mind, in this section, we opted to define our curvature differently from the simulation case in equation \eqref{eq:definition_kappa_simulation}. We will consider only the side of the sheet that displays the largest curvature in absolute value, while the opposite side is ignored since it is likely hiding an internal curvature that cannot be seen. Therefore, the curvature in this section will be defined as
\begin{equation}
\kappa_{max} =
\begin{cases}
\kappa_1, &\quad |\kappa_1| \geq |\kappa_2| \\
\kappa_2, &\quad \text{otherwise},
\end{cases}
\label{eq:definition_kappa_experiment}
\end{equation}
where $\kappa_1$ and $\kappa_2$ refer to the curvature of each side of the sheet as illustrated previously in figure \ref{fig:curv_circle_fit_method}.

We now sweep over the ratio between the diameters of the laser beam and the droplet $(d/D_0)$. This is done by keeping fixed the beam diameter $d = 20\mu  \text{m}$ and using droplets with diameter in the range $D_0 \in [27, \ 59]\mu \text{m}$. For each experiment, side view snapshots were taken at the times $[0, \ 500, \ 1000, \ 2000, \ 2500]\text{ns}$. The first two snapshots were used to estimate the expansion velocity of the sheet $\dot{R_0}$, which is then used to obtain the nondimensional expansion time $t_{exp} = {t \cdot \dot{R_0}/R_0}$. In order to measure the curvature at similar expansion times, we then select the snapshot of each experiment that has time closest to $t_{exp} = 5$. The actual selected time will vary between experiments since the experimental snapshots are only available at limited timesteps.

\begin{figure}[htbp]
\centering
\includegraphics[width=1.0\textwidth]{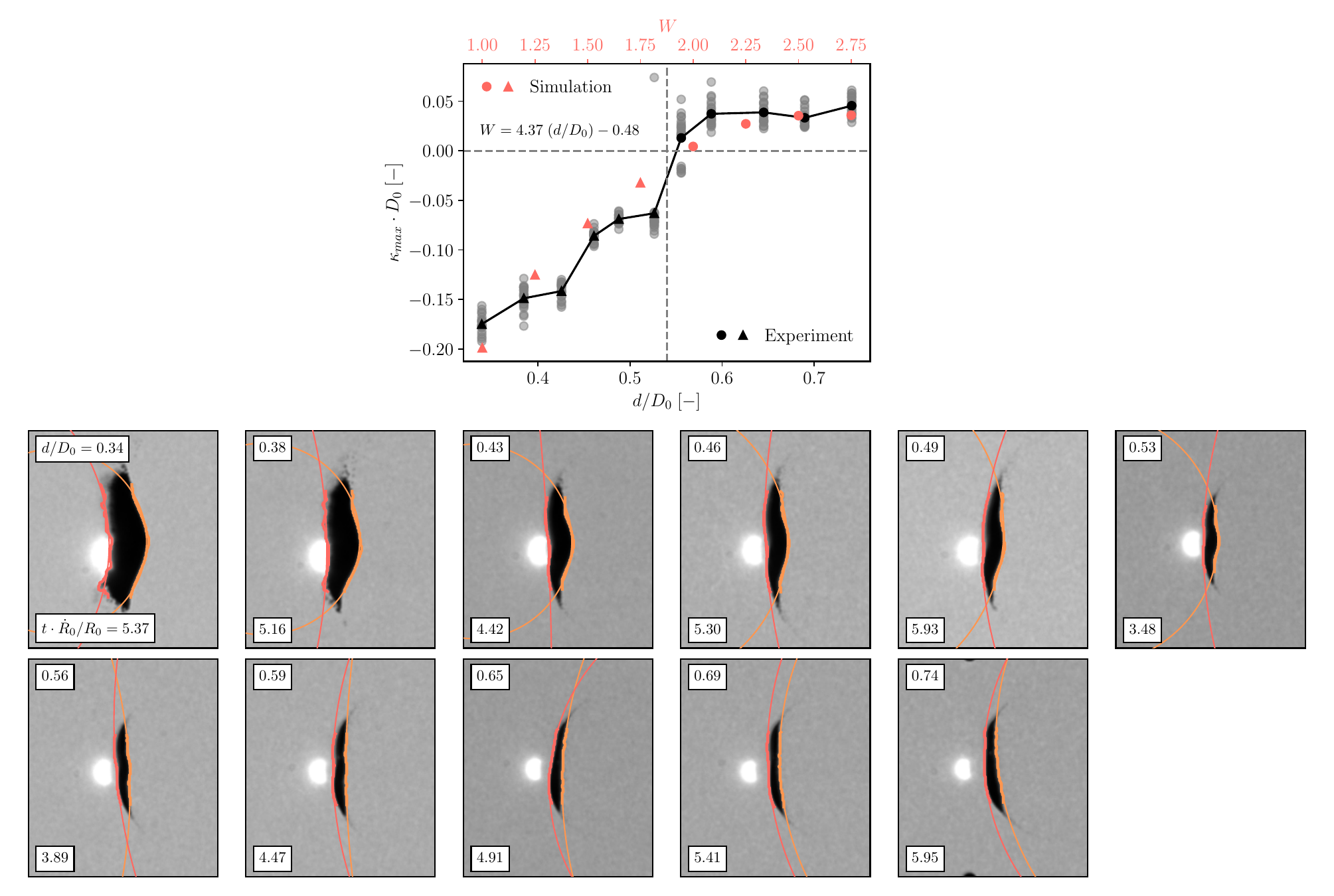}
\caption{Experimentally measured sheet curvature for different ratios $d/D_0$. For all cases shown, we use $E_{pp}=0.8$mJ, $d=20$um and $\tau_p=10$ns, while $D_0$ is varied from $27$um to $59$um. Top: the experimental curvature is plotted as a function of $d/D_0$ (black dots). As a qualitative comparison, the numerical curvature is shown as a function of the raised cosine parameter $W$ (red). Bottom: a shadowgraph for each $d/D_0$ is shown. The orange and red curves indicate the circles that were fitted to each side of the sheet, as described in figure \ref{fig:curv_circle_fit_method}. The bright spots are generated by plasma light that oversaturates the camera chip.}
\label{fig:experimental_kappa}
\end{figure}

Figure \ref{fig:experimental_kappa} shows the experimental images at the selected time. As expected, we observe that a focused laser pulse (with respect to the droplet size) pierces the center of the droplet, resulting in a sheet with negative curvature. On the other hand, a wide beam results in a sheet that curves positively. Since the droplet is small in comparison to the beam size, we hypothesize that the laser generates a plasma that wraps around most of the droplet, such that the tips will also experience a pressure that can push them forward. The value of $\kappa_{max}$ is plotted in panel a) as a function of $d/D_0$. The triangular points indicate the cases in which $\kappa_1$ was selected as $\kappa_{max}$, while for the circle points, $\kappa_2$ was selected. Similarly to the simulation results, the curvature flips from negative to positive as the ratio $d/D_0$ increases, eventually approaching a plateau. This flip from negative to positive curvature happens around $d / D_0 \approx 0.55$, indicated as a vertical line in figure \ref{fig:experimental_kappa}. For comparison, we also plot the simulated curvature as a function of $W$ in the same panel using a separate horizontal axis for $W$. The linear fit used to correlate the $W$ axis and the $d/D_0$ axis is given by $W = 4.37 \ (d/D_0) - 0.48$. This fit, however, is specific for the range of ratios shown here $d/D_0 \approx 0.5$ and would not realistically hold for more focused laser pulses, where a non-linear scaling would be necessary. We note that to fairly compare numerical and experimental results, the numerical curvature shown here is also $\kappa_{max}$, which is different from the curvature $\kappa_{avg}$ used in the previous sections. For simulations, we note that the flip between $\kappa_1$ or $\kappa_2$ being chosen for $\kappa_{max}$ happens at $W \approx 1.9$, which is also indicated by the vertical line in figure \ref{fig:experimental_kappa} that separates the triangular and circular simulation points. Overall, a good agreement between the experimental and simulated curvatures can be seen, which motivates the usage of the raised cosine parameter $W$ as a numerical analogous of the experimental ratio $d/D_0$.

\begin{figure}[htbp]
\centering
\includegraphics[width=1\textwidth]{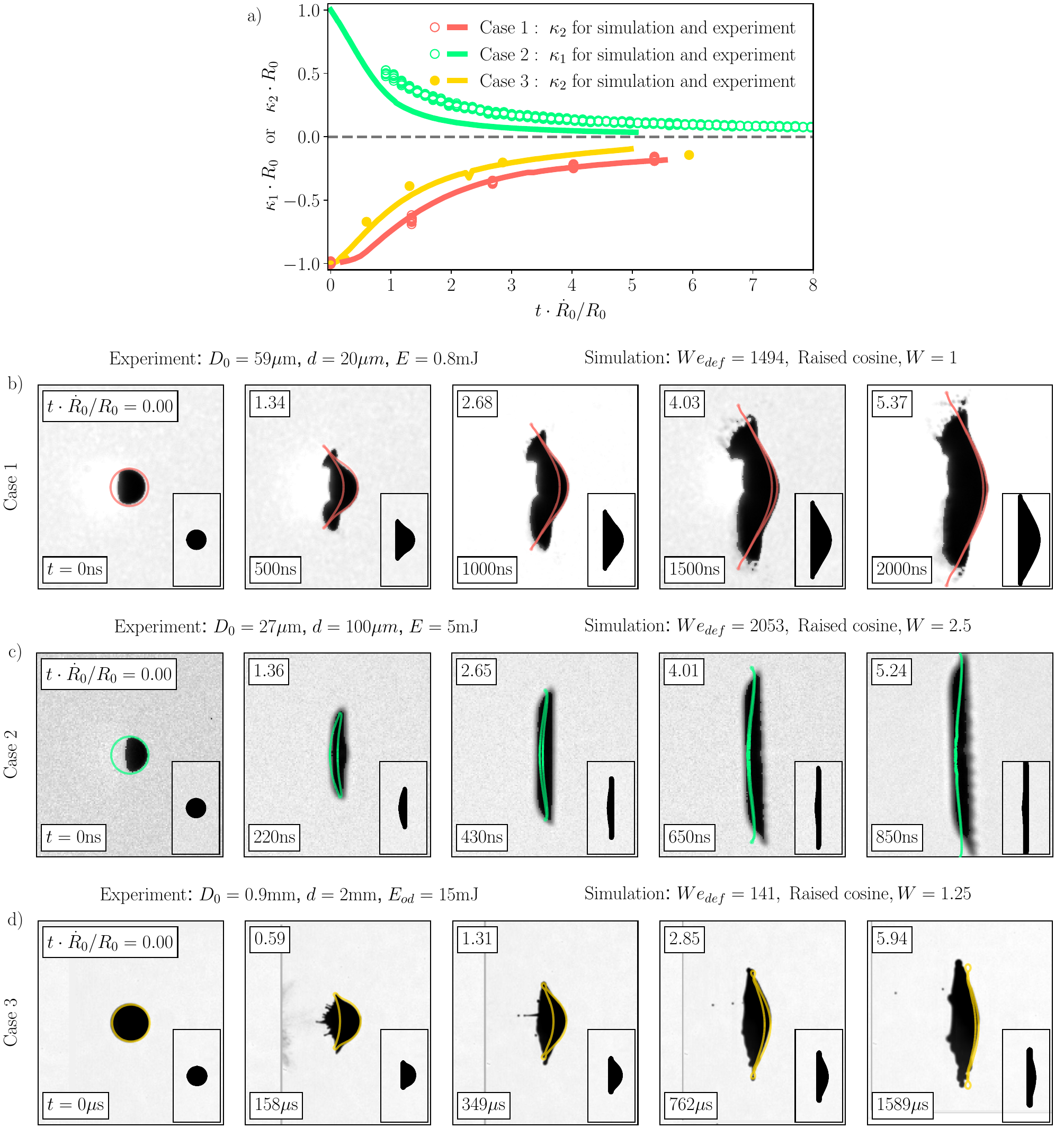}
\caption{Frame-by-frame comparison between sheets obtained experimentally and numerically for two different sets of parameters. The background of each plot shows the experimental side-view shadowgraph, while the red/green/yellow outlines show the simulated sheet cross-section. The insets show the side-view projection of the simulated sheet. Top row: a focused beam on a large droplet results in negative curvature. Panel a) Numerical and experimental curvature over time for each case in the following panels. Panel b) case from a focused laser beam resulting in negative curvature. Panel c) unfocused beam resulting in positive curvature. Panel d) experiment with a large water droplet reproduced from \cite{Klein2015}.}
\label{fig:visual_comparison_sim_exp}
\end{figure}

In order to visually compare the sheet deformation between experiments and simulations, in figure \ref{fig:visual_comparison_sim_exp} we show a timelapse of three different scenarios involving positive and negative curvatures. Each frame contains a shadowgraph of an experiment overlayed with the corresponding simulated sheet cross-section. The inset of each frame shows the projection of the three-dimensional simulated sheet, which would be the equivalent comparison to what is seen in the experimental images. The curvature $\kappa_{exp}$ is measured over time for all three cases and plotted at the top panel of the same figure.

Figure \ref{fig:visual_comparison_sim_exp}b contains shadowgraphs at five timestamps of a tin droplet experiment that exhibits negative curvature. To match this specific experiment, we perform a simulation with a focused raised cosine profile of $W = 1$ and an expansion Weber number of $We_{def} = 1494$. This overlayed comparison between simulation and experiment accentuates the main advantage of the numerical simulations: the actual sheet thickness can be seen in the simulation, while it is completely hidden in the projected 2D experimental images. Very good agreement is observed in the bulk area of the sheet between experiment and simulation, while some discrepancy is observed only as we get closer to the edges. Experimentally, strong asymmetric sheet fragmentation is observed at the edge of the sheets as a result of a violent expansion. As previously discussed, this behavior cannot be captured by our simulations, which are axisymmetric and only present the expected formation of a rim. In the edges of the experimental images, we also notice an area in which the sheet deviates from the bulk curvature, becoming nearly flat abruptly. This behavior is also not captured by our initialization approach, which creates a smooth curvature change along the droplet and sheet.

Figure \ref{fig:visual_comparison_sim_exp}c shows the same experimental-simulation visual comparison now for a case with positive curvature. We chose timestamps that result in expansion times similar to those in the previous case. At the earlier time of $t_{exp} = 1.36$ we observe a very good agreement between the simulated and experimental sheets. Note that no onset of experimental or numerical fragmentation has yet been observed at the sheet edge since the sheet volume is better distributed with a near-constant thickness profile in this case. At later times, while the positive curvature is maintained for both the simulated and experimental sheets, a discrepancy can be observed: the laser side of the sheet remains uniformly curved in the simulation, while in the experiments, it flattens out and only the edges curve forward. This is a striking feature of the experimental sheets that is still numerically unreproducible with our current choice of pressure profile. Numerically, we also see the formation of a small rim at later times that curves slightly back. The rim is formed due to capillary effects acting at a late time, and we believe this rim is slightly pulled back due to drag in the simulations (the outer medium is not a vacuum in simulations due to numerical limitations). 

Figure \ref{fig:visual_comparison_sim_exp}d showcases the laser-induced deformation of a larger water droplet, reproduced from the work of Klein et al. \cite{Klein2015}. In this experiment, a large water droplet with $D_0 = 0.9\text{mm}$ is used, and due to its size, the velocity received by the droplet is small, such that only a small Weber number is attained. We note that this is a striking difference from our experiments, in which a micrometer-sized droplet is used and deformed with a much higher Weber number $(We > 1000)$. Not only do the size, velocity, and time scales differ from our experiments, but also the mechanism of inducing the initial pressure. In this water droplet, only a vapor cloud is created, as opposed to the tin plasma that quickly expands and wraps the droplet in our experiments. Besides all these differences, we show that the raised cosine approach can also be successfully used to describe this flow. In this case, a focused profile with $W = 1.25$ was chosen to obtain the negative curvature observed in the experiment. The good visual agreement is observed in the frame-by-frame comparison, which is also confirmed in the $\kappa_{max}$ measurements over time shown in the top plot. This confirms that this approach can be used in a wide range of droplet impact scenarios, with different time, length, and velocity scales and different mechanisms of propulsion.

\section{Discussion, conclusion, and perspective}
\label{sec:conclusion}
We numerically and experimentally investigated the morphology of a liquid sheet obtained by laser-induced droplet deformation. Direct numerical simulations are performed to understand how the initial pressure profile exerted on the droplet surface can lead to different late-time sheet morphologies. We propose a pressure profile based on a raised cosine function and demonstrate that, by tuning its parameter $W$, we can obtain better morphological agreement with experiments compared to previous functions proposed in the literature.

Our results show that using an instantaneous pressure impulse, we can still obtain both negative and positive sheet curvatures as long as the pressure profile function is correctly chosen. The previously proposed profiles in literature (Gaussian and cosine functions) could not provide both sheet curvature types, which illustrates again the importance of carefully choosing or determining the pressure profile. The actual determination of this pressure profile from experiments is a difficult task since it comes from the complex interaction between a laser-generated plasma and a droplet. As this plasma expands and wraps the droplet, the determination of the exerted pressure is not trivial. Therefore, while we show in this work that we can obtain positive curvatures as long as a suitable pressure impulse is given, we note that more work on determining the real shape of this impulse from an experimental point of view is still required. Computationally, determining the ``correct'' pressure impulse would require complex plasma physics modeling of the interaction between tin plasma and the droplet, which requires different numerical techniques which are out of the scope of the current work.

We have analyzed the corresponding initial velocity field for a given pressure profiled proposed a method to predict whether a given profile will produce a positive or negatively curved sheet. This method shows that a positively curved sheet can be obtained with raised cosine functions of $W>2$, which can never be achieved with the traditional Gaussian pressure. We observe that in order to obtain positive curvature, a pressure profile needs to present a wide peak and a short tail, which the raised cosine can achieve due to its negative excess kurtosis but not the Gaussian (zero excess kurtosis). The initial prediction is then compared with the curvature obtained after the simulation of the droplet deformation. Good agreement is obtained as we see that, indeed, the sheet curves forward for approximately $W>2$. Predicting sheet morphology from the initial condition at time zero can be very beneficial for future works on probing different pressure profiles and their corresponding deformation dynamics. Without having to perform expensive dynamical fluid simulations, it is possible to test different profiles and tune them accordingly to obtain the desired morphology. This allows for efficient and careful design of pressure profiles to obtain different sheet morphologies.

Focusing on the raised cosine profile, we have shown that the numerical parameter $W$ can be correlated to the experimental beam-to-droplet size ratio $d/D_0$. A negative sheet curvature is obtained for low values of $W$ and $d/D_0$, while a positive curvature is obtained for high values. This effect is associated with the pressure profile being either focused at the center of the droplet or wrapping around most of the droplet surface, which can experimentally happen due to the fast plasma expansion.

The approach used in this work can be used more generally than only in laser-droplet applications for EUV light generation. We have shown that the raised cosine function also correctly captures the laser-induced expansion dynamics of a large millimeter-sized water droplet performed in \cite{Klein2015}. These deformation dynamics have also already been shown to be similar to the case of droplets impacting narrow solid pillars, such that this numerical simulation approach could also be used for droplet impact problems to design or optimize such systems.

\section*{Acknowledgements}
This work was conducted at the Advanced Research Center for Nanolithography (ARCNL), a public-private partnership between the University of Amsterdam (UvA), Vrije Universiteit Amsterdam (VU), Rijksuniversiteit Groningen (UG), the Dutch Research Council (NWO), and the semiconductor equipment manufacturer ASML and was partly financed by ‘Toeslag voor Topconsortia voor Kennis en Innovatie (TKI)’ from the Dutch Ministry of Economic Affairs and Climate Policy. 
\printbibliography

\appendix

\section*{Appendix: Numerical Validation}
\label{section:appendix_validation}
\begin{figure}[htbp]
\centering
\includegraphics[width=1\textwidth]{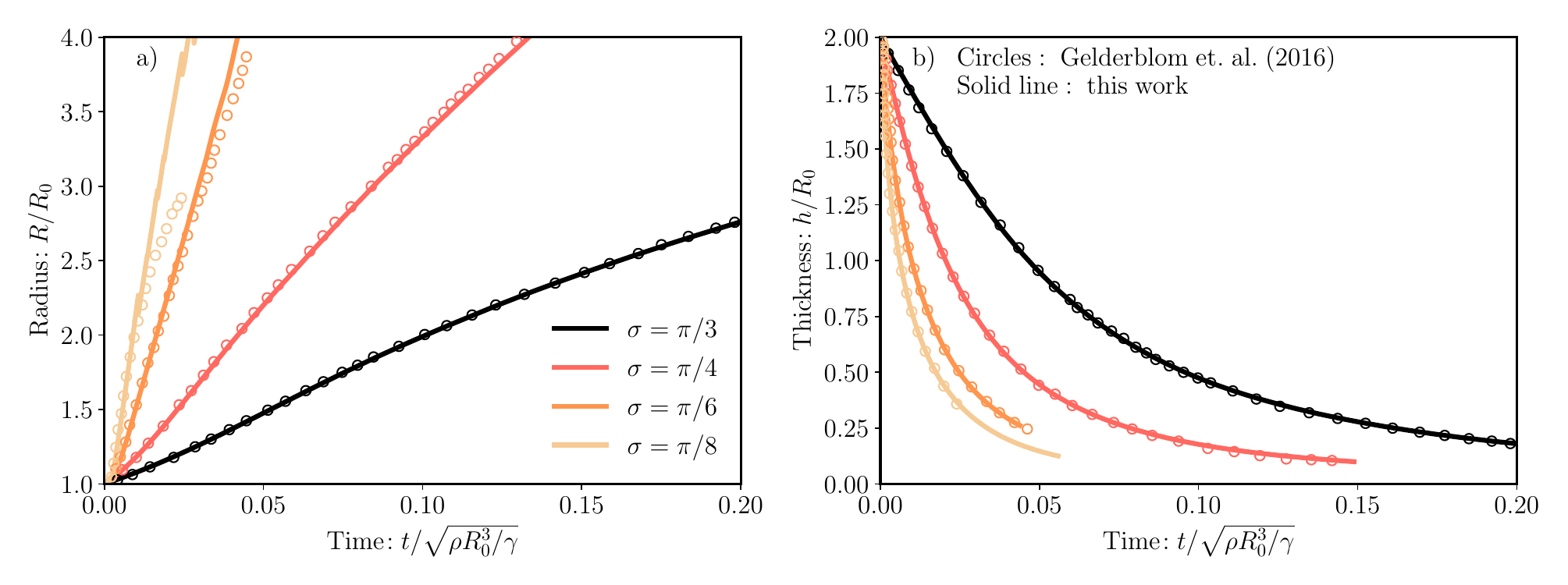}
\caption{Radius and thickness of the droplet/sheet over time for four different simulations. All simulations are performed with Gaussian pressure profiles, and the Gaussian width is varied in the set $\sigma \in \{ \pi/8, \ \pi/6, \ \pi/4, \ \pi/3 \}$. Results from this work (solid lines) are compared against the results from the Boundary Integral simulations in \cite{Gelderblom2016} (circles) for numerical validation of our implementation.}
\label{fig:hanneke_validation}
\end{figure}

To validate our implementation of this approach with Basilisk C, we compare our results against the Boundary Integral simulations performed in \cite{Gelderblom2016}. Figure \ref{fig:hanneke_validation} shows the radius and central thickness of the sheet over time for four simulations with different Gaussian pressure profiles. All four simulations were performed with fixed propulsion Reynolds and Weber numbers $Re = 5000$ and $We = 800$, respectively. The pressure profile was always kept as a Gaussian, and the Gaussian width was varied for each simulation within the set $\sigma \in \{ \pi/8, \ \pi/6, \ \pi/4, \ \pi/3 \}$. We notice that the curves for both the radius and the thickness agree well with the reference data from \cite{Gelderblom2016}, serving as validation of our solutions.

\section*{Appendix: Expansion-to-propulsion velocity ratio}
\label{section:velocity_ratio}
While the raised cosine profile can provide a more realistic direction for curvature, it still presents a discrepancy from experimental observations. This discrepancy comes in the form of the expansion-to-propulsion velocity ratio $\dot{R_0} / U_z$. This quantity indicates how fast the droplet expands compared to how fast it propels forward.

\begin{figure}[htbp]
\centering
\includegraphics[width=1\textwidth]{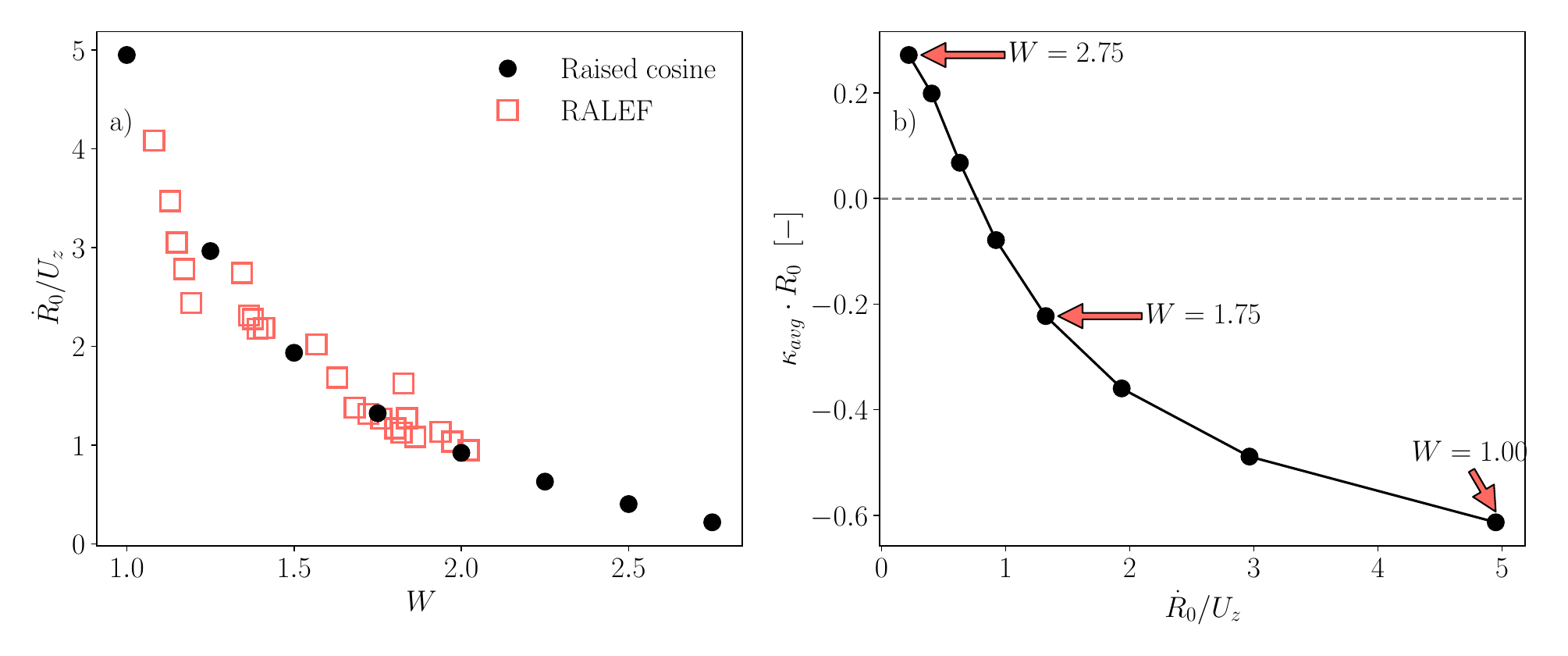}
\caption{Measurements of the velocity ratio $\dot{R}_0 / U_z$ for raised cosine profiles. a) Velocity ratio as a function of $W$ for the initialization approach in this work and as predicted by the code RALEF in \cite{HernandezRueda2022}. b) Numerical average curvature $\kappa_{avg}$ as a function of the velocity ratio in our simulations.  }
\label{fig:velocity_ratio}
\end{figure}

In figure \ref{fig:velocity_ratio}a, the blue circles show $\dot{R}_0/U_z$ as a function of the raised cosine width $W$. A very sharp diverging increase is observed for $W \rightarrow 0$, which indicates that almost all the energy is being directed to deforming the droplet and not to propel it. For higher values of $W$, the ratio decreases and eventually vanishes into a plateau where the droplet only propels and never expands. To validate if the raised cosine exhibits the expected velocity ratio for a given $W$, we compare our results with those previously reported by \cite{HernandezRueda2022} using the full radiation-hydrodynamic code RALEF-2D. In that software, the pressure profile is not given as an input. Instead, the actual laser parameters are provided, and the full laser interaction with the tin droplet is simulated, such that the pressure profile and $\dot{R}_0/U_z$ are both outputs of the simulation. We fitted the pressure profiles provided by RALEF in \cite{HernandezRueda2022} using raised cosines and plotted the velocity ratio provided by their code as a function of the fitted $W$. We can see that the velocity ratios obtained are a good match with the initialization approach used in our simulations, which validates the initialization method. 

As we have seen previously in figure \ref{fig:initial_deformed_curv_correlation}, in order to obtain significant positive curvature with a raised cosine, the width parameter needs to be $W > 2$. In this range of widths, we see in figure \ref{fig:velocity_ratio} that the velocity ratio is always below 1, such that we cannot obtain fast-expanding sheets that exhibit a positive curvature. This limitation is directly visible in figure \ref{fig:velocity_ratio}b, where we plot the predicted sheet curvature as a function of $\dot{R}_0/U_z$, and we can clearly see that we are only able to obtain simulations with positive curvature along with velocity ratios that are below 1. This is still a discrepancy from experiments that need to be further studied in the future, as somehow, the experimental sheets exhibit both positive curvature and fast expansion simultaneously.

\end{document}